\documentclass[twocolumn,showpacs,prb]{revtex4}

\usepackage{graphicx}

\begin{document}

\title{From Linear to Nonlinear Response in Spin
Glasses:\\Importance of Mean-Field-Theory Predictions}
\author{V.~S.~Zotev}
\author{R.~Orbach}
\affiliation{Department of Physics, University of California,
Riverside, California 92521}
\date{Submitted to PRB on December 27, 2001}

\begin{abstract}
Deviations from spin-glass linear response in a single crystal
Cu:Mn 1.5 at \% are studied for a wide range of changes in
magnetic field, $\Delta H$. Three quantities, the difference
$TRM-(MFC-ZFC)$, the effective waiting time, $t_{w}^{e\!f\!f}$,
and the difference $TRM(t_{w})-TRM(t_{w}=0)$ are examined in our
analysis. Three regimes of spin-glass behavior are observed as
$\Delta H$ increases. Lines in the $(T,\Delta H)$ plane,
corresponding to ``weak'' and ``strong'' violations of linear
response under a change in magnetic field, are shown to have the
same functional form as the de Almeida-Thouless critical line. Our
results demonstrate the existence of a fundamental link between
static and dynamic properties of spin glasses, predicted by the
mean-field theory of aging phenomena.
\end{abstract}


\maketitle

\section{\label{intro}Introduction}

The properties of spin glasses in a magnetic field have been the
subject of considerable attention. It is widely recognized that
the Parisi replica-symmetry breaking ansatz \cite{mez87} provides
an essentially correct mean-field solution for the infinite-range
Sherrington-Kirkpatrick model. \cite{she75} The spin-glass phase
in this model is separated from the paramagnetic phase in the
$(T,H)$ plane by the de Almeida-Thouless (AT) critical line.
\cite{alm78} The situation is less clear in the case of
finite-dimensional short-range models. Rigorous theoretical
results show viability of the mean-field approach for the
description of these systems, \cite{mar00}  but progress is
impeded by significant analytical difficulties. Numerical studies
have repeatedly suggested the existence of an AT-type critical
line at $d=3$ and higher dimensions. \cite{car90,zul98} However,
magnetic field effects present a difficult challenge for computer
simulations. Finite size, and difficulties in equilibrating large
samples, do not yet allow a clear distinction between the
mean-field picture and the droplet scenario. \cite{krz01} Because
of this, the existence of the spin-glass phase transition in a
magnetic field is considered a most relevant open
problem.\cite{zul98}

The experimental evidence for an AT-type critical behavior appears
ambiguous. Many real spin glasses have mean-field-like phase
diagrams, \cite{ken91} with an onset of strong $MFC-ZFC$
irreversibility along a certain line. This irreversibility is
usually interpreted as a sign of replica-symmetry breaking.
However, real spin glasses are always out of equilibrium, and the
measured AT lines are time dependent. Therefore, it appears
impossible to obtain information about the \emph{equilibrium}
phase diagram directly from experiments. We show in this paper,
however, that the study of magnetic field effects on the
\emph{nonequilibrium} dynamics of spin glasses can shed light on
the magnetic properties of the equilibrium spin-glass state.

Our motivation for this work is twofold. First, effects of
magnetic field are understood fairly well within the mean-field
theory, where they are derived from first principles. The minimum
possible overlap, $q_{min}(H)$, of two states in the presence of a
magnetic magnetic field, $H$, plays an important role in this
approach. Yet, as usually happens in spin-glass physics, the
theoretical predictions cannot be easily related to experimentally
observable phenomena. Even those models of spin-glass dynamics
that are based on the mean-field-like hierarchical picture of
phase space tend to treat magnetic field effects
phenomenologically in terms of the Zeeman energy. In this paper,
we show that experimental results on the magnetic field dependence
of spin-glass dynamics are consistent with the mean-field
description.

The second reason for this work is practical. Spin-glass
relaxation properties are usually studied under a change in
magnetic field. It is often believed that the subsequent response
is linear in magnetic field ``for reasonably small field values
(say $<10~G$)''. \cite{kop88} We show in this paper that the very
definition of what one means by ``reasonably small'' fields is
impossible without knowledge of the spin-glass phase diagram. This
knowledge becomes vital when results for different temperatures or
different samples are compared.

The paper is organized as follows. The next Section describes a
theoretical picture underlying our analysis. Sec.~III.A presents
results of magnetization measurements. Sec.~III.B is devoted to
experimental results for the effective waiting time. In
Sec.~III.C, the waiting-time dependence of the measured quantities
is discussed in detail. Sec.~IV summarizes our conclusions.

\section{\label{theor}Theoretical background}

\subsection{\label{theA} Mean-field dynamics}

The main obstacle to an experimental test of mean-field-theory
predictions is the problem of relating experimentally accessible
spin-glass dynamics to the static equilibrium properties of the
spin-glass state, described by the Parisi solution.
Phenomenological phase-space models have provided important
insights into possible physical mechanisms for spin-glass
relaxation.\cite{led91,bou92} A recent theoretical breakthrough
has been achieved within the mean-field theory of aging phenomena,
proposed by Cugliandolo \textit{et al}.
\cite{cug93,fra95,mar98,fra98} Because these theoretical
developments have profound significance for the interpretation of
our experimental results, we shall briefly review them here.

The well-known fluctuation-dissipation theorem (FDT) \cite{kub66}
establishes a link between the linear response of a system to an
external perturbation, and the fluctuation properties of the
system in thermal equilibrium. The response of a spin system at
time $t$ to an instantaneous field at time $t'$, $R(t,t')$, and
the spin autocorrelation function, $C(t,t')$,  are defined as
follows:
\begin{equation}
R(t,t')=(1/N)\sum_{i=1}^{N} \delta \langle S_{i}(t) \rangle /
\delta h(t')~~; \label{res}
\end{equation}
\begin{equation}
C(t,t')=(1/N)\sum_{i=1}^{N} \langle S_{i}(t)S_{i}(t') \rangle~~.
\label{cor}
\end{equation}
In thermodynamic equilibrium, both functions are time-translation
invariant, and related by the fluctuation-dissipation theorem
($\beta=1/k_{B}T$):
\begin{equation}
R_{eq}(t-t')= \beta \partial C_{eq} (t-t') / \partial t'~~.
\label{fdt}
\end{equation}
The violation of this theorem in a general off-equilibrium
situation is described by the function $X(t,t') \leq 1$:
\cite{cug93,fra95,mar98}
\begin{equation}
R(t,t')= \beta X(t,t') \partial C(t,t') / \partial t'~~.
\label{gen}
\end{equation}
Violation of the fluctuation-dissipation theorem is associated
with a deviation from linear response. A second-order nonlinearity
does not appear because the magnetization changes sign when the
field is reversed. It has been shown \cite{lev89} that the
presence of a third-order nonlinear response turns the
fluctuation-dissipation theorem into a fluctuation-dissipation
inequality. This is consistent with the fact that $X(t,t')$ in
Eq.~(4) is generally less than unity.

When these formulas are applied to spin-glass dynamics, the
relevant times are $t=t_{w}+\tau$ and $t'=t_{w}$, where $t_{w}$
and $\tau$ are the waiting time and observation time after the
waiting time, respectively. The dynamical quantities $C$ and $X$
can be related to their equilibrium counterparts if both $t_{w}$
and $\tau$ are sent to infinity. However, time-translational
invariance does not hold off equilibrium, and the limiting values
depend on how the limits are taken. One can write for the
correlation function, \cite{bou97,cug99}
\begin{equation}
\lim_{\tau\rightarrow\infty} \lim_{t_{w}\rightarrow\infty}
C(t_{w}+\tau, t_{w}) = q_{EA}~~, \label{qea}
\end{equation}
and
\begin{equation}
\lim_{\tau\rightarrow\infty} C(t_{w}+\tau, t_{w})=q_{min}~~.
\label{qmi}
\end{equation}
Eq.~(5) is a dynamical definition of the Edwards-Anderson order
parameter, $q_{EA}$, as the value of the correlation function at
the limit of validity of the fluctuation-dissipation theorem.
Eq.~(6) expresses the property of weak ergodicity breaking in spin
glasses. If the waiting time is finite and the magnetic field is
zero (giving $q_{min}=0$), the system is able to escape
arbitrarily far from the configuration it had reached at
$t=t_{w}$. \cite{bou97}

The fluctuation-dissipation ratio $X(q)$ is defined as the limit
of the function $X(t,t')$, taken along a path in the $(t,t')$
plane, characterized by a given value $q$ of the correlation
function $C(t,t')$.\cite{fra98} The most important result of the
theory is the fact that, under the assumption of stochastic
stability,\cite{fra98} $X(q)$ is equal to the equilibrium order
parameter function $x(q)$:
\begin{equation}
X(q)=\lim_{t,t'\rightarrow\infty;~C=q}
\frac{R(t,t')}{\beta\partial C(t,t')/\partial t'}=x(q)~~.
\label{fdr}
\end{equation}
The function $x(q)$ is an integral of the Parisi order parameter
$P(q)$,\cite{mez87} which can be introduced for short-range spin
glasses within the standard replica-symmetry breaking formalism.
\cite{mar00}  According to Eq.~(7), there is a deep relationship
between spin-glass dynamics and the static equilibrium properties
of the spin-glass state. It implies that ``in any
finite-dimensional system, replica-symmetry breaking and aging in
the response functions either appear together or do not appear at
all''. \cite{fra98}

We now consider the physical (experimental) significance of these
theoretical results. It is well known that there are two main
regimes of spin-glass relaxation. In the equilibrium, or
stationary regime, observation times $\tau \ll t_{w}$, the
fluctuation-dissipation theorem holds and the response function
depends only on $\tau$. In the nonequilibrium, or aging regime,
observation times $\tau > t_{w}$, the FDT is violated and the
relaxation depends on $t_{w}$ for any $\tau$. Both regimes have
been observed experimentally \cite{oci85} and studied
numerically.\cite{fra95,and92} The transition from one regime to
the other is marked by a peak in the relaxation rate,
corresponding to strong violation of FDT at $\tau \approx t_{w}$.

These effects can be naturally explained within the mean-field
theory of aging phenomena. It has been suggested \cite{cug93} that
the function $X(t,t')$, introduced to describe violations of the
FDT in Eq.~(4), depends on its time arguments only through the
correlation function: $X=X[C(t,t')]$. In the equilibrium regime,
the correlation function decreases rapidly (on the linear time
scale) from 1 to $q_{EA}$, and $X(C)=1$. In the aging regime, the
correlation function relaxes slowly from $q_{EA}$ to $q_{min}$,
and $X(C)<1$. \cite{cug99,par98} The transition from one regime to
the other is a direct consequence of replica-symmetry breaking at
$q<q_{EA}$.

\subsection{\label{theB} Chaotic nature of the spin-glass state}

Another important issue is the chaotic nature of the spin glass
state with respect to magnetic field. It has been demonstrated
numerically that a small change in external field leads to a
considerable reorganization of a spin configuration.\cite{ban81}
The Parisi solution suggests that an average equilibrium overlap
between two states at different but similar magnetic fields,
$(h_{1}-h_{2})^{2} N \gg 1$,\cite{mar00} is equal to $q_{min}$,
i.e. the minimum possible overlap.\cite{par83} Analysis of
fluctuations around the Parisi solution, carried out by
Kondor,\cite{kon89} demonstrates that the correlation overlap
function $C_{H}(r)$ for two spins, $i$ and $j$, at a distance $r$,
behaves as
\begin{equation}
C_{H}(r)= \overline{ \langle S_{i} S_{j} \rangle_{H} \langle S_{i}
S_{j} \rangle_{0}} ~ \propto ~ \exp(-r/\xi_{H})~. \label{chr}
\end{equation}
This means that the projection of the correlation $\langle
S_{i}S_{j} \rangle_{H}$ at field $H$ onto the correlation $\langle
S_{i}S_{j} \rangle_{0}$ at zero field vanishes beyond the finite
characteristic length $\xi_{H}$. Near $T_{g}$ and at low fields,
this magnetic correlation length is simply $\xi_{H}=1/q_{min}(H)$.
In the mean-field theory, $q_{min}(H) \propto H^{2/3}$, so that
$\xi_{H}$ diverges rapidly as the field goes to zero.

The behavior of the correlation function, Eq.~(8), is a
consequence of replica-symmetry breaking. The low-temperature
spin-glass phase has an essentially infinite number of pure
equilibrium states. Each of them is characterized by an infinite
correlation length. These states have equal free energies per
site, except for differences of the order $O(1/N)$. Only a few
states with the lowest energies contribute significantly to the
partition function. Because of the small energy differences, any
small (but finite) amount of energy, added to the system, is
enough to reshuffle the Boltzmann weights of the different states
and thus completely reorganize the equilibrium spin configuration.
Application of a magnetic field is an example of such a
perturbation.

The magnetic correlation length, $\xi_{H}$, has the following
meaning.\cite{rit94} When a magnetic field $H$ is applied to the
system, the minimum possible overlap of two states is equal to
$q_{min}(H)$. Consequently, all the states having overlaps $q <
q_{min}$ are suppressed by the field. Their free energies
increase, and they acquire the finite correlation length
$\xi_{H}$. The spatial spin correlations, corresponding to these
states, survive only within this range. All the other pure states
are still characterized by an infinite correlation length. At the
AT line, where $q_{min}=q_{EA}$, all the states have a finite
correlation length, and the system becomes paramagnetic. Thus, a
change in magnetic field has a randomizing effect on the
spin-glass state. The ratio $q_{min}/q_{EA}$ is a natural measure
of this effect.

The chaotic nature of the spin-glass state is also reflected in
the phenomenological droplet model. The magnetic correlation
length is determined as an average droplet size for which the
Zeeman energy is equal to the energy of the droplet excitation. It
is given by the following expression:\cite{fis88}
\begin{equation}
\xi_{H} \propto H^{-2/(d-2\theta)}~. \label{xih}
\end{equation}
Even though this result, with $d=3$ and $\theta \approx 0.2$, is
similar to the result of the mean-field theory, the physics is
quite different. Spin-glass properties in this model are governed
by low-energy excitations of the ground state, created by coherent
flipping of compact clusters of spins. It is suggested that the
magnetic field, $H$, would flip all the droplets with sizes
greater than $\xi_{H}$ and thus destroy \emph{all} spin-glass
correlations beyond this length scale. Therefore, at any nonzero
field, there is only a paramagnetic state with the finite
correlation length $\xi_{H}$.

In both these approaches, the chaotic nature of the spin-glass
state with respect to magnetic field leads to departures from
linear response as $\Delta H$ increases.

\section{\label{exper}Experimental results and analysis}

The purpose of this paper is the study of violations of the
fluctuation-dissipation theorem as a function of magnetic field
change $\Delta H$. According to Eq.~(7), aging dynamics of the
spin-glass state is ultimately determined by static equilibrium
properties. Any violation of the FDT contains information about
replica-symmetry breaking. Therefore, studies of the gradual
deviation from linear response as $\Delta H$ increases can provide
insight into the nature of the spin-glass phase diagram. Direct
experimental determination of the correlation function, Eq.~(2),
requires sophisticated measurements of the time dependent magnetic
noise spectrum.\cite{bou88}  Instead, we make use of quantities
which can be obtained from magnetic susceptibility measurements.
The first quantity of interest is the difference of the values of
remanence, measured in TRM and ZFC experiments: $TRM-(MFC-ZFC)$.
It has been shown \cite{lun86} that, within the linear regime,
this quantity is zero, provided that all three magnetizations are
measured at the same time after the initial quench. The second
quantity of interest is the effective waiting time,
$t_{w}^{e\!f\!f}$, defined as the value of the observation time
$\tau$ where the relaxation rate $S(\tau)=-\partial
TRM(\tau,t_{w},\Delta H)/ \partial \log \tau $ has a maximum. It
has been argued \cite{dju95} that, within the linear response
regime, $t_{w}^{e\!f\!f}\approx t_{w}$, so that the peak in
$S(\tau)$ is essentially unaffected by a small magnetic field
change. The third quantity we study experimentally is the
difference $TRM(t_{w})-TRM(t_{w}=0)$. It describes a change in the
measured magnetization as a result of the waiting time, and thus
allows closer examination of aging phenomena.

All experiments were performed on a single crystal of Cu:Mn 1.5 at
\%, a typical Heisenberg spin glass with a glass temperature of
about $15.2~K$. Results of conventional studies of the phase
diagram for this sample will be reported elsewhere. A commercial
Quantum Design SQUID magnetometer was used for all the
measurements.

\subsection{\label{expA} Measurements of $TRM$, $MFC$, and $ZFC$ }

Fig.~1 exhibits the dependence of $TRM$, $MFC-ZFC$, and their
difference, on the field change $\Delta H$, measured at
$T=12.0~K$. All data points were taken at the same short
observation time, $\tau \approx 40~s$, with zero waiting time, and
an effective cooling time of about $600~s$. Error bars are smaller
than the symbol sizes. The same will apply to all figures without
error bars. Fig.~1 shows that three different types of spin-glass
behavior can be distinguished for different ranges of magnetic
field change, $\Delta H$. For field changes from $\Delta H=0$ to
$\Delta H_{1}\approx 120~ Oe$, the difference $TRM-(MFC-ZFC)$ is
zero to within our experimental accuracy. Between $\Delta H_{1}$
and $\Delta H_{2}\approx 420~Oe$, there are weak deviations from
linear response. At $\Delta H=\Delta H_{2}$, corresponding
approximately to the peak in the $MFC-ZFC$ remanence, violation of
linear response becomes strong, and the difference $TRM-(MFC-ZFC)$
is a linear function of $\Delta H$ with a large slope.

\begin{figure}
\resizebox{\columnwidth}{!}{\includegraphics{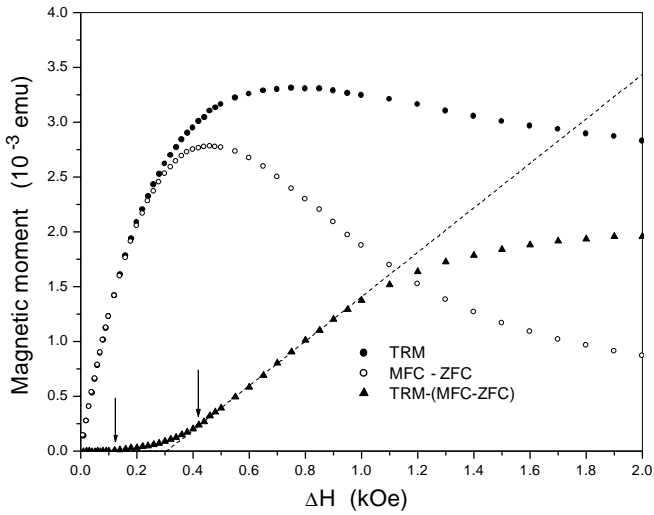}}
\caption{\label{Fig1} The values of the remanent magnetic moment,
measured at $T=12.0~K=0.79~T_{g}$ and $t_{w}=0$. The magnetic
field is changed from $H=\Delta H$ to $H=0$ (TRM experiment) and
from $0$ to $\Delta H$ (ZFC experiment). The two arrows indicate
the critical field changes, $\Delta H_{1} \approx 120~Oe$ and
$\Delta H_{2} \approx 420~Oe$, for the weak and strong linear
response violations, respectively.}
\end{figure}

The temperature dependences of the critical field changes $\Delta
H_{1}$ and $\Delta H_{2}$ are presented in Fig.~2 and Fig.~3. The
figures also show the critical AT line, determined for the same
sample at the same observation and cooling times, using the onset
of strong $MFC-ZFC$ irreversibility as the signature of the
spin-glass phase transition. We assume that this dynamical line
approximates the equilibrium AT line for infinite waiting time.
The difference $MFC-ZFC$ in Fig.~1 is not zero at the AT field of
$\approx 1400~Oe$ because transverse freezing above the AT line
leads to a weak longitudinal irreversibility.\cite{gab81}

\begin{figure}
\resizebox{\columnwidth}{!}{\includegraphics{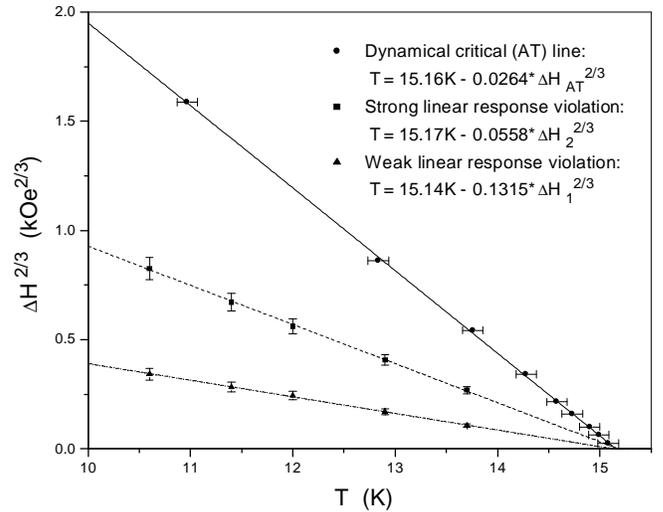}}
\caption{\label{Fig2} The dynamical AT line $\Delta H_{AT}(T)$,
the strong linear response violation line $\Delta H_{2}(T)$, and
the weak linear response violation line $\Delta H_{1}(T)$,
determined from the linear fits to experimental data, plotted as
$\Delta H^{2/3}$ vs. $T$.}
\end{figure}

\begin{figure}
\resizebox{\columnwidth}{!}{\includegraphics{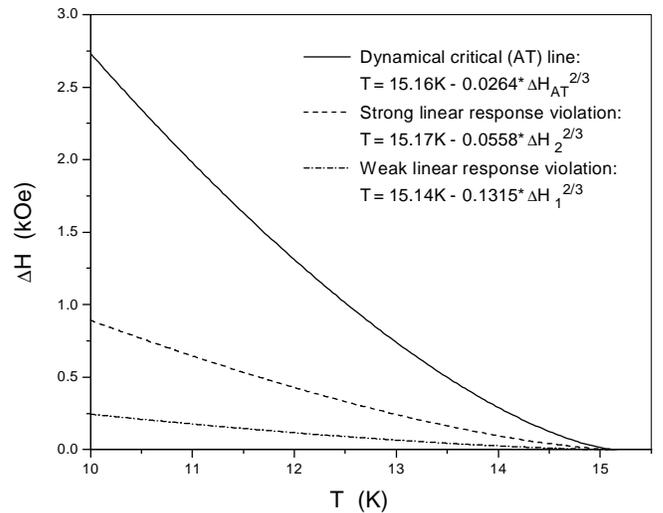}}
\caption{\label{Fig3} The same lines as in Fig.~2, but plotted as
$\Delta H(T)$. Note that the two linear response violation lines
below the AT line represent a dynamic crossover from one
relaxation regime to the other. They are \emph{not} features of an
equilibrium phase diagram.}
\end{figure}

One sees from Fig.~2 that all three lines in the $(T,\Delta H)$
plane have the same functional form, $T_{g}-T=a\Delta H^{2/3}$,
typical of the equilibrium AT critical line \cite{alm78} in the SK
model. Moreover, the two linear-response violation lines, $\Delta
H_{1}(T)$ and $\Delta H_{2}(T)$, determine dynamical transition
temperatures which are very close to the actual glass temperature
$T_{g}$. It would thus be incorrect to say that the critical
AT-type line plays a role at high fields only. Our results
demonstrate that this line manifests itself dynamically even at
very low fields. A critical field change for a given degree of
linear response violation seems to be a constant fraction of the
AT field. It means, from a practical point of view, that results
obtained at different temperatures can be directly compared only
if they have the same $\Delta H / H_{AT}(T)$ ratio.

A comment should be made at this point. It has been argued
\cite{wen84} that a $T \propto H^{2/3}$ dependence for transition
lines is not a unique feature of the mean-field theory. A similar
power law could be obtained from purely dynamical considerations
without using the concept of the spin-glass phase
transition.\cite{wen84} A dynamical freezing line,
$T_{f}(H,\omega)$, appears in the droplet model as
well.\cite{fis88} In the present analysis, we do not rely on the
existence of the $H^{2/3}$ dependence by itself. Our argument is
based on the fact that the experimental AT and linear response
violation lines have the \emph{same} functional form, as
illustrated in Fig.~2 and Fig.~3. This result suggests that there
is a close relationship between the static and dynamic properties
of the spin-glass state, assuming that the measured AT line is
related to the equilibrium one.

Of course, measured values of the critical field changes depend on
both the waiting time and the observation time. This point is
illustrated in Fig.~4 where values of $TRM-(MFC-ZFC)$ are plotted
for $t_{w}=0$ and $t_{w}=30~min$. As the waiting time becomes
larger, $\Delta H_{1}$ increases, while $\Delta H_{2}$ decreases.
Thus, longer equilibration times make the system less susceptible
to external perturbations, but only up to a certain point. This
issue will be discussed further in Sec.~III.C.
\begin{figure}
\resizebox{\columnwidth}{!}{\includegraphics{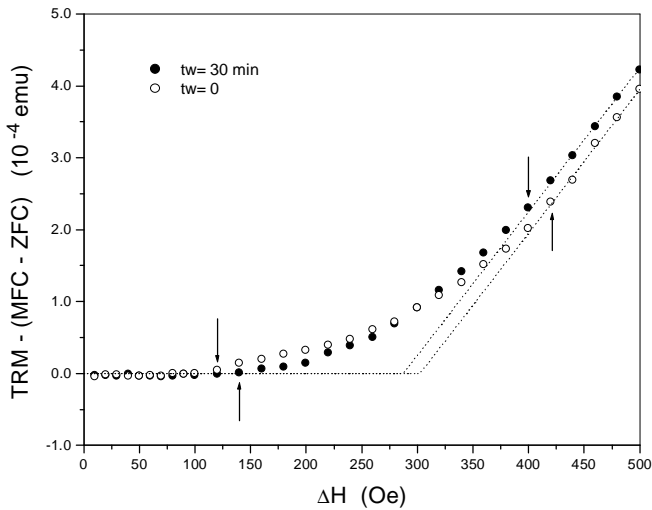}}
\caption{\label{Fig4} The function $TRM-(MFC-ZFC)$, describing
violation of linear response, at $t_{w}=30~min$ and $t_{w}=0$. The
second and the third arrows from the left indicate, respectively,
the larger value of $\Delta H_{1}$ and the lower value of $\Delta
H_{2}$ for the longer waiting time.}
\end{figure}

\subsection{\label{expB} Measurements of the effective waiting time }

Let us now turn to a discussion of the effective waiting time,
$t_{w}^{e\!f\!f}$. Fig.~5 exhibits the dependence of
$\log_{10}(t_{w}^{e\!f\!f})$ on the magnetic field change $\Delta
H$ for three different waiting times: $t_{w}=30~min,~5~min$, and
0. The temperature is the same as in Fig.~1. One can easily
discern three regimes of FDT violation, corresponding to the three
regimes of deviation from linear response seen in Fig.~1.
\begin{figure}
\resizebox{\columnwidth}{!}{\includegraphics{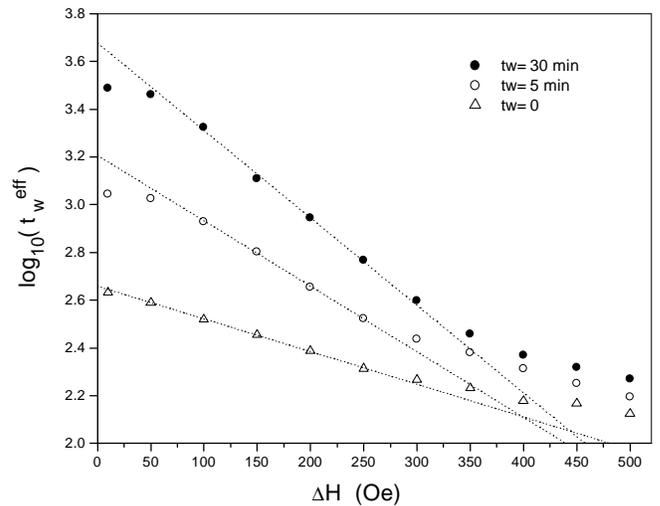}}
\caption{\label{Fig5} Logarithm of the effective waiting time,
$t_{w}^{e\!f\!f}$, as a function of $\Delta H$, measured at
$T=12.0~K=0.79~Tg$ for three waiting times. The dotted lines are
linear fits in the interval $100...250~Oe$. They are only guides
for the eye.}
\end{figure}

As the magnetic field change, $\Delta H$, increases from zero up
to about $100~Oe$, the effective waiting time decreases only
slightly. This means that FDT holds for observation times less
than $t_{w}^{e\!f\!f} \approx t_{w}$. Of course, if the waiting
time is short, $t_{w}^{e\!f\!f}$ is determined by the cooling
procedure.

In the interval from $100~Oe$ to about $400~Oe$, the effective
waiting time drops sharply, and the field dependence of its
logarithm is linear. In this case, nonequilibrium behavior appears
much earlier, and the FDT holds only at short times $\tau \ll
t_{w}^{e\!f\!f} \ll t_{w}$. This corresponds to the weak deviation
from linear response between $\Delta H_{1}$ and $\Delta H_{2}$ in
Fig.~1.

When the field change exceeds $400~Oe$, the FDT is strongly
violated. One can see from Fig.~5 that, in this regime, the curves
for long waiting times approach the curve for zero waiting time,
and the waiting time dependence almost disappears. This situation
corresponds to the strong deviation from linear response at short
observation times above $\Delta H_{2}$ in Fig.~1. Therefore, the
break from the linear dependence of $\log_{10}(t_{w}^{e\!f\!f})$
on $\Delta H$, observed at large field changes, is directly
related to strong nonlinearity in the spin-glass response. The
effective waiting time in this case is determined primarily by the
zero waiting time results, dependent on the experimental protocol.

The waiting time dependence in Fig.~5 supports the conclusion,
drawn from Fig.~4, that an increase in waiting time leads to an
expansion of the linear response region to higher values of
$\Delta H$. An increase in the observation time $\tau$ has the
opposite effect. The field dependence is more pronounced at longer
times. This accounts for the fact that the critical field changes,
extracted from the effective waiting time experiments (Fig.~5),
seem to be lower than those obtained from the short-time
magnetization measurements (Fig.~4). Another interesting feature
of our results is that the dependence of
$\log_{10}(t_{w}^{e\!f\!f})$ on $\Delta H$ for $t_{w}=0$,
determined exclusively by the cooling procedure, does not exhibit
the low-field plateau seen at longer waiting times. This behavior
is analyzed in Sec.~III.C.

The temperature dependence of the effective waiting time is
exhibited in Fig.~6 and Fig.~7, where $\log_{10}(t_{w}^{e\!f\!f})$
vs. $\Delta H$ data for $t_{w}=30~min$ are plotted for four
different temperatures. The results for the smallest field change
of $10~Oe$ are not exactly the same because the effective cooling
time increases as the measurement temperature is lowered. It is
quite evident from Fig.~7, however, that the four curves scale
rather well together if plotted vs. $\Delta H / \Delta H_{2}$.
This means that the critical AT line sets a characteristic
magnetic field scale at any temperature below $T_{g}$, and that it
plays an important role for \emph{all} aspects of spin-glass
dynamics.

\begin{figure}
\resizebox{\columnwidth}{!}{\includegraphics{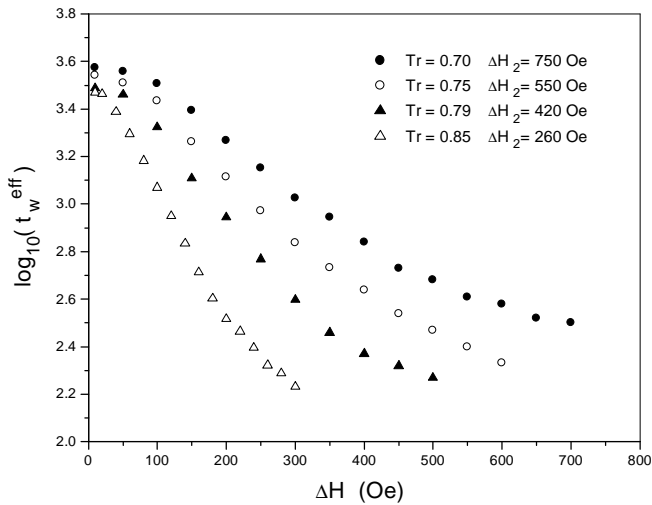}}
\caption{\label{Fig6} Logarithm of the effective waiting time,
$t_{w}^{e\!f\!f}$, measured as a function of $\Delta H$ for
$t_{w}=30~min$ at four temperatures: $Tr=T/T_{g}=0.70,~0.75,~0.79$
and $0.85$. Note the similarity in shape and pronounced difference
in the magnetic field scales.}
\end{figure}

\begin{figure}
\resizebox{\columnwidth}{!}{\includegraphics{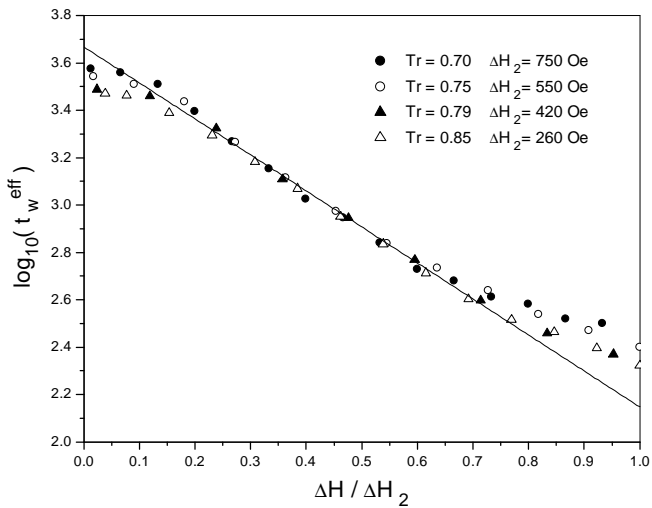}}
\caption{\label{Fig7} The experimental results of Fig.~6, plotted
vs. $\Delta H/\Delta H_{2}$. The critical field change $\Delta
H_{2}$, corresponding to strong violation of linear response, is
proportional to the AT field, $H_{AT}(T)$. The straight line is a
guide for the eye.}
\end{figure}

\subsection{\label{expC} Waiting time dependence of $\Delta H_{1}$ and $\Delta H_{2}$ }

The waiting time dependences of the characteristic field changes,
corresponding to the weak and strong linear response violations,
deserve special attention. Our experimental results, exhibited in
Fig.~4 and Fig.~5, suggest that $\Delta H_{1}$ increases with
$t_{w}$, while $\Delta H_{2}$ diminishes.

In order to check this conclusion, we have measured the field
dependence of the thermoremanent magnetization,
$TRM(\tau,t_{w},\Delta H)$, for three different waiting times:
$t_{w}=30~min,~5~min$, and 0. The measurement temperature is
$T=12.0~K$, and the observation time after the field change is
$\tau=40~s$. For convenience, we shall refer to the
zero-waiting-time dependence as $ZTRM(\tau,\Delta H)$. This
nonequilibrium relaxation function is determined by the cooling
process and not by the waiting time. The results are exhibited in
Fig.~8. One can see that the maxima in the TRM curves shift
towards lower fields as the waiting time increases. To examine
evolution of the measured magnetization curves due to aging
phenomena, we consider the differences $TRM-ZTRM$, where both
$TRM$ and $ZTRM$ are measured at the same observation time. Fig.~8
suggests that the maxima in $TRM-ZTRM$ correspond to higher fields
at longer waiting times.

\begin{figure}
\resizebox{\columnwidth}{!}{\includegraphics{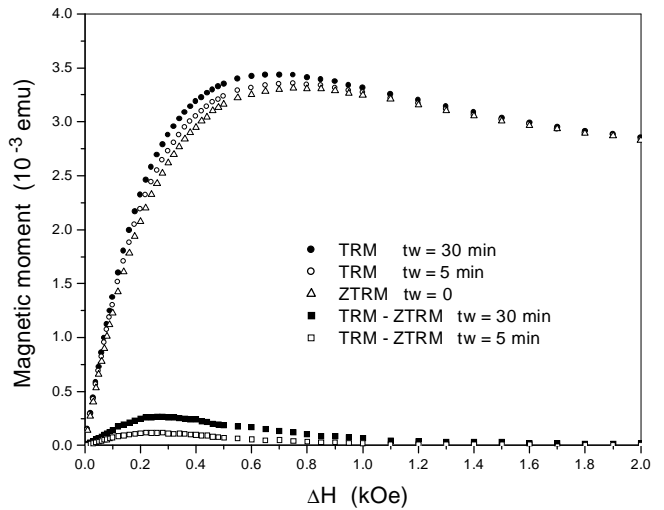}}
\caption{\label{Fig8} Comparison of $TRM(\Delta H)$ curves
measured at $T=12.0~K=0.79~T_{g}$ for different waiting times. The
curve for $t_{w}=0$ is referred to as $ZTRM$. As $t_{w}$
increases, the peak in $TRM$ shifts to the left, while the peak in
$TRM-ZTRM$ moves to the right.}
\end{figure}

Fig.~9 exhibits the logarithm of the ratio $(TRM-ZTRM)/\Delta H$.
Two different types of behavior can be clearly distinguished. The
ratio increases as long as the field change is less than about
$120~Oe$, and decreases at higher $\Delta H$. The field dependence
of its logarithm can be approximated by straight lines in both
regimes. The results in Fig.~9 can be understood if we compare
them with the data in Fig.~5. At low field changes, the effective
waiting time for $ZTRM$ decreases more steeply with increasing
$\Delta H$ than the effective waiting time for $TRM$, which has a
plateau in this region. The decay of the $ZTRM$ becomes faster in
time, and the ratio $(TRM-ZTRM)/\Delta H$, measured at fixed
observation time, increases with $\Delta H$. At higher magnetic
field changes the drop in the effective waiting time for the $TRM$
is greater than for the $ZTRM$. As a result, the ratio
$(TRM-ZTRM)/\Delta H$ decreases. According to Fig.~9, the
transition from one regime to the other occurs at a higher
magnetic field change for a longer waiting time. This observation
supports our conclusion that the plateau in the field dependence
of the effective waiting time broadens as the system equilibrates.

\begin{figure}
\resizebox{\columnwidth}{!}{\includegraphics{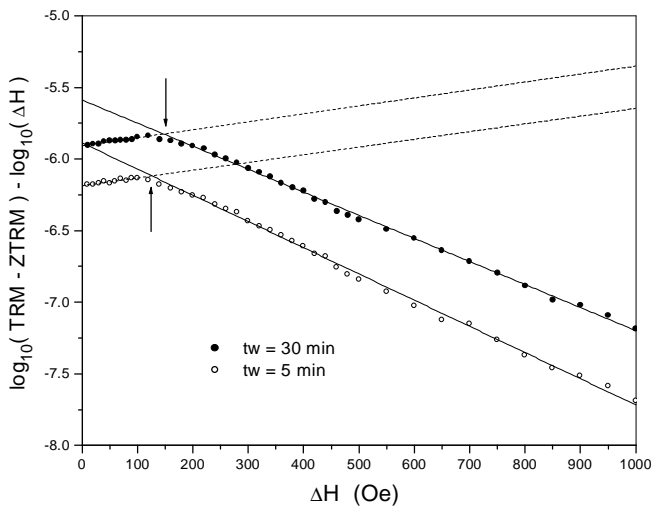}}
\caption{\label{Fig9} Logarithm of the normalized increase in the
thermoremanent magnetization, $(TRM-ZTRM)/\Delta H$, as a function
of $\Delta H$ at a constant observation time $\tau=40~s$. The
dashed lines are linear fits in the interval $0...100~Oe$. The
solid lines are linear fits in the interval $200...1000~Oe$. The
arrows indicate points where nonlinearity appears.}
\end{figure}

\begin{figure}
\resizebox{\columnwidth}{!}{\includegraphics{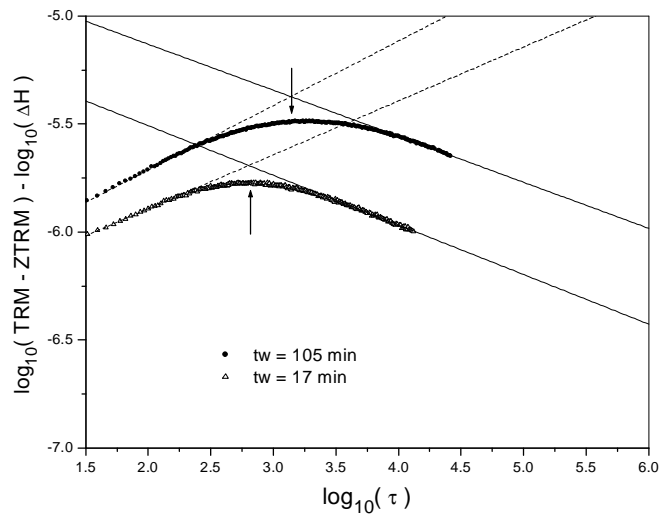}}
\caption{\label{Fig10} Logarithm of the normalized increase in the
thermoremanent magnetization, $(TRM-ZTRM)/\Delta H$, as a function
of $\log(\tau)$ at a constant field change $\Delta H=10~Oe$. The
straight lines are guides for the eye. The arrows indicate
transitional points.}
\end{figure}

Fig.~10 provides further insight into the nature of this effect.
It displays the same quantity, the logarithm of $(TRM-ZTRM)/\Delta
H$, as a function of time at constant $\Delta H=10~Oe$. The
experimental protocol is the same as before, except for the
different waiting times. One can see a deep similarity between the
results in Fig.~9 and Fig.~10. The waiting time dependence
exhibited in Fig.~10 is readily understandable: the system stays
longer in the quasiequilibrium regime for longer $t_{w}$, so the
maximum in $TRM-ZTRM$ shifts towards longer observation times.
Comparison with Fig.~9 suggests that the same is true for the
field dependence: the system in the quasiequilibrium regime can
sustain a stronger perturbation for a longer waiting time.

This similarity between the effects of the observation time and of
the field change on the spin-glass state is a natural consequence
of replica-symmetry breaking. The weak ergodicity breaking
scenario, mentioned in Sec.~II.A, suggests that, as the
observation time increases, the system can evolve very far from
the state at time $t_{w}$, with the minimum correlation set by
$q_{min}$. This results from the fact that the average relaxation
time is infinite because of the exponential distribution of free
energies of different states. Chaotic nature of the spin-glass
state with respect to magnetic field, discussed in Sec.~II.B,
forces the new equilibrium configuration after a field change to
have the minimum overlap $q_{min}$ with the old configuration.
This phenomenon is also related to existence of an essentially
infinite number of states with very similar free energies. Thus,
both an increasing observation time and an increasing field change
make the spin-glass state less correlated with the state at time
$t_{w}$.

It is of interest to examine the waiting time dependences of
$\Delta H_{1}$ and $\Delta H_{2}$ for the two ``competing''
descriptions of spin-glass dynamics: the droplet model and the
phase-space picture.

In the droplet model,\cite{fis88} if $R(t_{w})$ is an average
droplet size, and $L_{\tau}$ is an observation length scale, two
limiting cases can be distinguished. For $R(t_{w}) \ll \xi_{H}$,
the FDT is violated when $L_{\tau} \sim R(t_{w})$, and the field
does not play a significant role. For $\xi_{H} \ll R(t_{w})$, the
FDT is violated as soon as $L_{\tau} \sim \xi_{H}$, and the
waiting time is relatively unimportant. These regimes correspond
approximately to experimental regimes with $\Delta H<\Delta H_{1}$
and $\Delta H>\Delta H_{2}$, as discussed above. There is also an
intermediate regime with $R(t_{w}) \sim \xi_{H}$. This regime is
very interesting physically, because violation of the FDT depends
on interplay among all three length scales: $L_{\tau} \sim
R(t_{w}) \sim \xi_{H}$. Unfortunately, no predictions for this
regime are given within the droplet model.\cite{fis88} It has been
argued\cite{kop88} that, if the crossover between linear and
nonlinear regimes is defined by a condition $R(t_{w})=\xi_{H}$,
the field change, needed to provoke nonlinear relaxation, should
decrease with increasing $t_{w}$. This argument can explain the
waiting time dependence of $\Delta H_{2}$. However, it fails in
the case of $\Delta H_{1}$, which \emph{increases} with $t_{w}$.
Therefore, the experimentally observed waiting time dependence of
$\Delta H_{1}$ appears rather counterintuitive within the droplet
scenario.

The phase-space picture of spin-glass dynamics does provide a
consistent explanation for these phenomena. In this approach,
evolution of a system in the phase space can be viewed as a series
of transitions among traps, separated by free-energy barriers.
\cite{bou92} As the system equilibrates, it encounters traps with
higher barriers, and these traps increasingly resemble the pure
states, contributing to equilibrium.\cite{cug93} The
quasiequilibrium relaxation regime at short observation times
corresponds to evolution within each trap, and the subsequent
nonequilibrium behavior is related to evolution from trap to trap.
As the waiting time, $t_{w}$, increases, the system has to
overcome a higher effective barrier to leave a trap. This takes a
longer time, or (if the observation time is fixed) a larger field
change. Therefore, the characteristic field change for the weak
linear response violation, $\Delta H_{1}$, increases with $t_{w}$.
However, after overcoming a higher barrier, the system can explore
a broader free-energy landscape due to hierarchical structure of
the phase space. It takes less time or a smaller field change to
produce a state very different from the one at $t_{w}$. The
nonequilibrium behavior is thus more affected by external
perturbations for longer waiting times. Therefore, the
characteristic field change for the strong linear response
violation, $\Delta H_{2}$, decreases with $t_{w}$.

The transition from the quasiequilibrium to the nonequilibrium
regime is better defined after longer waiting times, both as a
function of the observation time and as a function of the field
change. This conclusion agrees with results of numerical
simulations, which show the same effect when susceptibility is
studied as a function of correlation.\cite{fra95}
\\

\section{\label{concl}Summary}

The usual method for studying the spin-glass phase diagram is the
observation of the $MFC-ZFC$ irreversibility as a function of
temperature at a fixed magnetic field, $H= \Delta H$. A different
approach is employed in the present paper. The temperature is kept
constant and `` irreversibility of the irreversibility'',
$TRM-(MFC-ZFC)$, is measured as a function of magnetic field
change $\Delta H$. Increasing $\Delta H$ leads to a gradual
deviation from linear response, and, equivalently, to a violation
of the fluctuation-dissipation theorem. This violation, according
to the mean-field theory of aging phenomena, is directly related
to replica-symmetry breaking. Our experiments show that both
methods give the same functional form for the critical lines,
suggesting a mean-field-like phase diagram. This conclusion is
further supported by measurements of the effective waiting time,
$t_{w}^{e\!f\!f}$. The study of waiting time effects on the linear
response violation suggests the validity of the phase-space
picture for spin-glass dynamics. Our results demonstrate the
existence of a fundamental link between static and dynamic
properties of spin glasses, predicted by the mean-field theory of
aging phenomena.\\

We would like to thank Professor~J.~A.~Mydosh for providing us
with the Cu:Mn single crystal sample, prepared in Kamerlingh Onnes
Laboratory (Leiden, The Netherlands). We are also grateful to
Dr.~G.~G.~Kenning for numerous interesting discussions and to
Dr.~J.~Hammann and Dr.~E.~Vincent from CEA Saclay (France) for
help and advice.

\end{document}